\documentclass{Interspeech}

\interspeechcameraready

\title{Towards One-bit ASR: Extremely Low-bit Conformer Quantization \\Using Co-training and Stochastic
Precision}

\author[affiliation={1}]{Zhaoqing}{Li}
\author[affiliation={1}]{Haoning}{Xu}
\author[affiliation={2}]{Zengrui}{Jin}
\author[affiliation={1}]{Lingwei}{Meng}
\author[affiliation={1}]{Tianzi}{Wang}
\author[affiliation={1}]{Huimeng}{Wang}
\author[affiliation={1}]{Youjun}{Chen}
\author[affiliation={1}]{Mingyu}{Cui}
\author[affiliation={1}]{Shujie}{Hu}
\author[affiliation={1}]{Xunying}{Liu}

\affiliation{}{The Chinese University of Hong Kong}{China}
\affiliation{Department of Electronic Engineering}{Tsinghua University}{China}
\email{zqli@se.cuhk.edu.hk, xyliu@se.cuhk.edu.hk}

\usepackage{comment}

\usepackage{cite}
\usepackage{amsmath,amssymb,amsfonts}
\usepackage{algorithmic}
\usepackage{graphicx}
\usepackage{textcomp}
\usepackage{xcolor}

\usepackage{booktabs}
\usepackage{color,multirow,float,enumitem,colortbl}
\usepackage{bibspacing}
\usepackage{bbding}
\usepackage{pifont}
\newcommand{\cmark}{\ding{51}}
\newcommand{\xmark}{\ding{55}}

\begin{document}

\maketitle

\begin{abstract}
    Model compression has become an emerging need as the sizes of modern speech systems rapidly increase. In this paper, we study model weight quantization, which directly reduces the memory footprint to accommodate computationally resource-constrained applications. We propose novel approaches to perform extremely low-bit (i.e., 2-bit and 1-bit) quantization of Conformer automatic speech recognition systems using multiple precision model co-training, stochastic precision, and tensor-wise learnable scaling factors to alleviate quantization incurred performance loss. The proposed methods can achieve performance-lossless 2-bit and 1-bit quantization of Conformer ASR systems trained with the 300-hr Switchboard and 960-hr LibriSpeech corpus. Maximum overall performance-lossless compression ratios of 16.2 and 16.6 times are achieved without a statistically significant increase in the word error rate (WER) over the full precision baseline systems, respectively.
\end{abstract}

\section{Introduction}
Modern automatic speech recognition (ASR) models such as Conformer have achieved significant progress on various speech recognition scenarios~\cite{gulati2020conformer,peng2022branchformer,baevski2020wav2vec,hsu2021hubert,yao2023zipformer,watanabe2017hybrid}. However, performance progress tends to accompany an increasing number of model parameters and the need for computation and storage resources~\cite{yeh2022efficient}. As a result, model compression with a neutral or even negligible quality impact becomes an important research topic. 

In this paper, we focus on model weights quantization, a straightforward and effective compression technique, that replaces floating-point stored weight parameters with quantized integers~\cite{gholami2022survey}. Particularly, extremely low-bit quantization (e.g., lower than 4 bits) can further benefit the resource-intensive deployment cases, as it further reduces the model footprint and speeds up the calculation to some extent (e.g., a binary model containing values of only -1 and 1 can perform addition operations instead of multiplications). This motivates recent 2-bit and 1-bit quantization studies in computer vision~\cite{hubara2016binarized,rastegari2016xnor,lin2017towards,liu2018bi,choi2019accurate} and language~\cite{wang2023bitnet,ma2024era,liu2022bit,bai2021binarybert} and speech~\cite{xiang2017binary,qian2019binary} processing tasks. Despite the success of many works on quantization of ASR models with no less than 4 bits~\cite{ding20224,kim2022integer,li2023lossless,fasoli20214,xu2021mixed,li2024one,xu2021mixed2}, extremely low-bit quantization techniques have not yet been successfully demonstrated for ASR models, as significant performance degradation can be observed in terms of the speech recognition word error rate (WER) when those ASR systems are quantized to 2 bits or 1 bit. For example, \cite{ding2024usm} reports a WER increase of over 200\% (relative) when using quantization with 2 bits. Also, \cite{yeh2022efficient} reports WER increases of more than 100\% (relative) when the model is quantized to 2 bits or less. Therefore, this paper studies the challenging 2-bit and most challenging 1-bit quantization of Conformer ASR systems.


Few recent efforts have been made to study the quantization of end-to-end (E2E) ASR models with lower than 4 bits~\cite{li2023lossless,rybakov20232,rybakov2024usm,yuancompressed}. Despite the progress achieved by these studies, they suffer from one or more of the following limitations: \textbf{1) significant performance degradation}~\cite{li2023lossless,rybakov20232,rybakov2024usm,yuancompressed}. For example, relative increases in WER of more than 10\% and 15\% are reported in~\cite{yuancompressed} for 2-bit and 1-bit quantization, respectively. \cite{rybakov20232} obtained a 2-bit Conformer model at the cost of 17\% relative WER degradation compared to the 4-bit version with large-scale training data. In addition, the absence of statistical significance tests in~\cite{rybakov20232,rybakov2024usm,yuancompressed} makes it difficult to qualitatively assess whether a \textit{performance-lossless} quantization is achieved in terms of the final model performance. \textbf{2) a large number of extra quantization parameters} that increase the memory footprint and complexity of model deployments. The sub-channel quantization technique adopted in~\cite{rybakov20232,rybakov2024usm} splits a weight matrix into many sub-blocks (e.g., 64 blocks in~\cite{rybakov2024usm}) and then quantizes each block with independent quantization parameters. \textbf{3) multi-stage operations} which are complicated or are dedicated to specific circumstances. \cite{yuancompressed} first distill an MoE-Conformer to a standard dense Conformer and then quantize the distilled Conformer. \cite{li2023lossless} first quantize a Conformer model to three sub-models of 8 bits, 4 bits, and 2 bits separately, and then search for a mixed-precision configuration of a target bit width by measuring the quantization error sensitivity with the three quantized systems.

To this end, this paper proposes a novel quantization-aware training framework for extremely low-bit (i.e., 2-bit or 1-bit) quantization of state-of-the-art Conformer speech recognition systems. We progressively adopt several methods to achieve a \textit{performance-lossless} quantization under ultra-low bit, including learnable scaling factors, quantization co-training, KL-divergence regularization, and stochastic precision. Specifically, we use tensor-wise learnable scaling factors that contain only negligible amounts of quantization parameters. We design a quantization co-training framework to allow simultaneous training of various weight-sharing low-bit systems. Based on this framework, we further use the KL-divergence regularization to enable explicit guidance (i.e., distillation) from higher-precision systems to lower-precision ones, which can be conducted concurrently within the training cycle and avoids multi-stage operations. To better benefit from distillation, we propose a stochastic precision approach to further reduce the performance gap between the teacher and student systems. The proposed methods are easy to implement and are flexible and extendable to many other quantization techniques such as sub-channel quantization.
The main contributions of the paper are summarized below:
\begin{itemize}[leftmargin=*]
    \item We present a quantization co-training framework, which allows simultaneously training a 2-bit model and using it to guide a 1-bit sub-model in a one-pass training manner. 
    \item We propose a novel stochastic precision technique, which samples partial layers of 2-bit model and binarizes them to 1-bit in each training iteration. The sampled sub-model intends to serve as an intermediate transition between 2-bit and 1-bit models, which can further reduce the performance gap of 2-bit and 1-bit models against the full-precision system. 
    \item We conduct intensive experiments to demonstrate the superior compression performance of our methods, compared to both state-of-the-art quantization methods and full-precision systems on the Switchboard and LibriSpeech datasets.
\end{itemize}


\vspace{-0.2cm}
\section{Conformer ASR Systems}
\vspace{-0.1cm}
The convolution-augmented Transformer (a.k.a. Conformer) is a popular E2E ASR architecture that achieves state-of-the-art performance on many speech recognition tasks~\cite{gulati2020conformer}. Similar to Transformers, a Conformer encoder consists of multiple blocks stacked together, where each block is further composed of the following modules in sequence: a feed-forward module (FFN), a self-attention module (MHSA), a convolution module (Conv), and a second FFN module (macaron-like) at the end. Among these modules, the FFN and MHSA modules account for most of the encoder parameters.

It is typically an effective way to train an attention-based encoder-decoder (AED) Conformer model by using the loss composed of multitask criterion interpolation~\cite{watanabe2017hybrid} between the CTC and attention of a Decoder (i.e., a smaller Transformer), which can be given by

\begin{equation}
\setlength\abovedisplayskip{0cm}
\setlength\belowdisplayskip{0cm}
    \mathcal{L}_{conformer}=(1-\gamma)\mathcal{L}_{att}+\gamma\mathcal{L}_{ctc},
    \label{eq1}
\end{equation}
where $\gamma$ is a constant and empirically set as 0.2
in this paper.

\begin{figure}[t]
    \centering
    \includegraphics[scale=0.35]{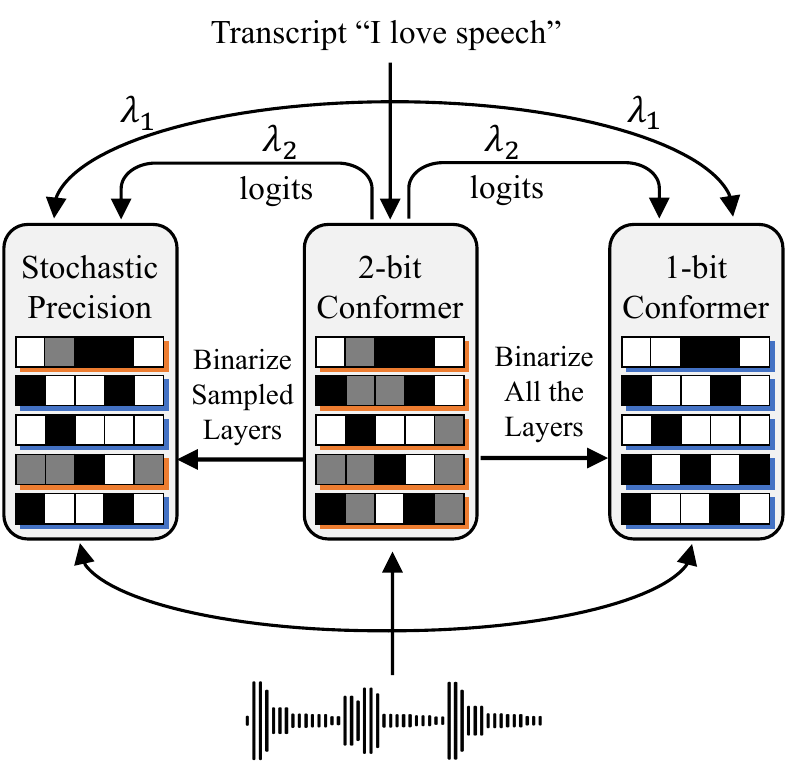}
    \vspace{-0.3cm}
    \caption{Diagram of quantization co-training framework with stochastic precision. When training, for each utterance, the 1-bit model is binarized from the 2-bit quantized model. The stochastic precision model is sampled by randomly binarizing partial layers of the 2-bit model. The three models share weight parameters. The training signal for 1-bit or stochastic precision model mixes the model loss and the KL-divergence regularization, which are calculated from real transcripts and the logits of the 2-bit model, respectively.}
    \label{fig1}
    \vspace{-0.6cm}
\end{figure}

\vspace{-0.2cm}
\section{Quantization Aware Co-Training}
\vspace{-0.1cm}
In this paper, we focus specifically on extremely low-bit quantization of Conformer ASR systems. In particular, our goal is to develop Conformer models with 2-bit or 1-bit weight parameters, which can reduce the memory footprint and also provide speed-up due to lower memory usage~\cite{rybakov20232}. To do so, we propose a new quantization co-training framework as illustrated in Fig.~\ref{fig1}. Then, we progressively introduce each of the adopted techniques.

\vspace{-0.25cm}
\subsection{Quantization-Aware Training with Learnable Scales}
\vspace{-0.15cm}
Model quantization replaces floating point stored weight parameters with low-precision integers. An $n$-bit quantization of a weight matrix $\mathcal{W}$ can be denoted as:
\vspace{-0.cm}
\begin{equation}
    \setlength\abovedisplayskip{0cm}
    \setlength\belowdisplayskip{0cm}
    \hat{\mathcal{W}}=\alpha\times\Pi_{\mathcal{Q}_n}(clip(\mathcal{W}/\alpha,min(\mathcal{Q}_n), max(\mathcal{Q}_n))),
    \label{eq2}
\end{equation}

\vspace{-0.cm}

\noindent where $\mathcal{W}$ and $\hat{\mathcal{W}}$ are full-precision and quantized model weights, respectively. $\Pi_{\mathcal{Q}_n}(\cdot)$ is a projection function that projects each element of $\mathcal{W}$ to its closest value in a quantization table $\mathcal{Q}_n$. In this paper, we focus on symmetric quantization. Specifically, we have $\mathcal{Q}_2=\alpha\times\{-1,0,+1\}$ and $\mathcal{Q}_1=\alpha\times\{-1,+1\}$ for 2-bit and 1-bit quantization, respectively, which are also referred to as ternary and binary cases. $\alpha\in\mathbb{R}$ is the tensor-wise scaling factor to control the quantization range of the entire weight matrix $\mathcal{W}$.

We use the Straight-Through Estimator (STE)~\cite{bengio2013estimating} to propagate back the gradient through the non-differentiable quantization function. For 1-bit quantization, unlike previous studies that compute the \textit{absmean} values of each channel or sub-channel as $\alpha$, we make $\alpha$ also a trainable parameter to better reduce the quantization error. More importantly, instead of assigning each channel a scaling factor, we consider tensor-wise scaling factors for each parameter tensor, which only needs negligible quantization parameters. Specifically, we update $\alpha$ by gradient descent method with the gradient given by\footnote{In Eq.~\ref{eq3}, the $sign(\cdot)$ is derived by scaling the actual gradient by $max(|\mathcal{Q}_n|)$ to make the training process more stable.}:

\begin{equation}
\setlength\abovedisplayskip{0cm}
\setlength\belowdisplayskip{0cm}
    \resizebox{.9\hsize}{!}{$
    \frac{\partial\hat{\mathcal{W}}}{\partial\alpha}=\left\{\begin{aligned}
        &-\mathcal{W}/\alpha + \Pi_{\mathcal{Q}_n}(\mathcal{W}/\alpha)& &\operatorname{if}|\mathcal{W}/\alpha| < max(|\mathcal{Q}_n|)\\
        &sign(\mathcal{W}/\alpha)& &\operatorname{if}|\mathcal{W}/\alpha| \geq max(|\mathcal{Q}_n|)
    \end{aligned}
    \right..
    \label{eq3}
    $}
\end{equation}

\vspace{-0.25cm}
\subsection{Quantization Co-Training}
\vspace{-0.15cm}
Co-training is a training scheme that enables simultaneous training of various weight-sharing sub-models (i.e., models comprise partial layers of the largest one). Experimental results in~\cite{touvron2023co}  show that co-training can improve the performance of sub-models against individually trained ones for computer vision tasks. In this paper, we also consider applying the co-training method to improve the performance of quantized models. However, we achieve this with a different point of view: a quantized model with lower precision can be regarded as a sub-model of one with higher precision since the former can be obtained by dropping partial bit-width of the latter. As such, we introduce a quantization co-training framework jointly training weight-sharing 2-bit and 1-bit quantized Conformer systems to improve their performance. The training loss is given by

\begin{equation}
    \setlength\abovedisplayskip{0cm}
    \setlength\belowdisplayskip{0cm}
    \mathcal{L}=\mathcal{L}_{int2}+\lambda_1\mathcal{L}_{int1},
    \label{eq4}
\end{equation}
where $\mathcal{L}_{int2}$ and $\mathcal{L}_{int1}$ are the loss of 2-bit and 1-bit quantized sub-models, respectively. $\lambda_1$ is a constant coefficient.

Actually, the co-training framework has further advantages. On one hand, it only needs to store the 2-bit model, while the 1-bit model is automatically contained as they share weight parameters, which saves cost for storing multiple systems of different precisions. On the other hand, when simultaneously training multiple systems, larger models can concurrently provide guidance to smaller ones within the training cycle, avoiding multi-stage operations as our next design.

\vspace{-0.25cm}
\subsection{KL-Divergence Regularization}
\vspace{-0.15cm}
Similar to the efficacy of model distillation, guiding low-precision models with high-precision teacher models can improve their performance. However, if a teacher is too much better than a student model, it may lead to sub-optimal performance for the student since the performance gap between them is too large, hindering the knowledge transfer. Therefore, in this paper, we use a 2-bit model to guide the 1-bit system to ensure a small performance gap. It is worth noticing that we can efficiently achieve this by incorporating the distillation process into the training cycle, eliminating the need to progressively perform multi-stage distillations that first train a 2-bit model and then distill a 1-bit model with it, as the practice in~\cite{liu2022bit,kim2019qkd}.

With our co-training framework, this is achieved by additionally introducing a regularization term. Formally, denoting $p_{int2}$ and $p_{int1}$ as the output distribution of 2-bit and 1-bit model, respectively, to guide a 1-bit student, we compute the regularization term with the Kullback–Leibler (KL) divergence between $p_{int2}$ and $p_{int1}$, given by
\begin{equation}
    \Omega_{int1} = D_{KL}(SG(p_{int2})||p_{int1}),
    \label{eq5}
\end{equation}
where $SG(\cdot)$ denotes the stop-gradient operation preventing the gradient from flowing back to the 2-bit teacher. Then, we can write the criterion with KL regularization:

\begin{equation}
    \mathcal{L}=\mathcal{L}_{int2}+\lambda_1\mathcal{L}_{int1}+\lambda_2\Omega_{int1},
    \label{eq6}
\end{equation}
where $\lambda_2$ is a constant coefficient.

\vspace{-0.1cm}
\subsection{Stochastic Precision}
\vspace{-0.1cm}
Although a 2-bit teacher is better than a full-precision one in terms of providing guidance for a 1-bit student, there still exist performance gaps that may lead to sub-optimal performance for both 2-bit and 1-bit models. This is because, in extremely low-bit cases, when model redundancy is already extremely squeezed out, model performance becomes more sensitive to quantization error. 

To address this problem, we propose a stochastic precision method that serves as an intermediate transition between the 2-bit and 1-bit models. Specifically, in each training iteration, we additionally train a sub-model with stochastic precision (i.e., between 2 bits and 1 bit) that is sampled by randomly binarizing partial layers of the 2-bit model. We also let the 2-bit model guide the stochastic-precision model. Finally, we have the following training criterion:
\begin{equation}
    \setlength\abovedisplayskip{0.2cm}
    \setlength\belowdisplayskip{0.1cm}
    \mathcal{L}=\mathcal{L}_{int2}+\lambda_1(\mathcal{L}_{int1}+\mathcal{L}_{SP})+\lambda_2(\Omega_{int1}+\Omega_{SP}),
    \label{eq7}
\end{equation}
where $\mathcal{L}_{SP}$ and $\Omega_{SP}$ are the sub-model loss and KL regularization term of the stochastic-precision model, respectively. The ultimate design incorporating all the techniques previously mentioned is illustrated in Fig.~\ref{fig1}.

\vspace{-0.2cm}
\section{Experiments}
\vspace{-0.1cm}
\subsection{Experimental Setup}
\vspace{-0.1cm}
We conduct experiments on two commonly used ASR datasets: 1) training on the 300-hr Switchboard corpus~\cite{godfrey1992switchboard} and testing on NIST Hub5’00, RT02, and RT03 test sets. 2) training on the LibriSpeech 960-hr datasets~\cite{panayotov2015librispeech} and testing on test-clean, test-other, dev-clean, and dev-other testsets. The full-precision baseline Conformer systems are configured using the ESPnet~\cite{watanabe2018espnet} recipes in ``espnet/egs2".
All the quantized systems are trained with the same configuration as their baselines. In this work, we mainly focus on low-bit quantization of the encoder, which accounts for the major part of the model parameters, and the Decoder is quantized with 4 bits unless otherwise stated. For stochastic-precision sub-models, in each iteration, the encoder layers are binarized at probabilities increasing from 0.2 to 0.9 for layers from 1 to 12 with a log-linear schedule, similar to a layer drop schedule. $\lambda_1$ and $\lambda_2$ are set to 0.5 and 1.0, respectively, which are to balance the losses of different components according to their respective magnitudes.

\vspace{-0.2cm}
\begin{figure}[h]
    \centering
    \includegraphics[scale=0.35]{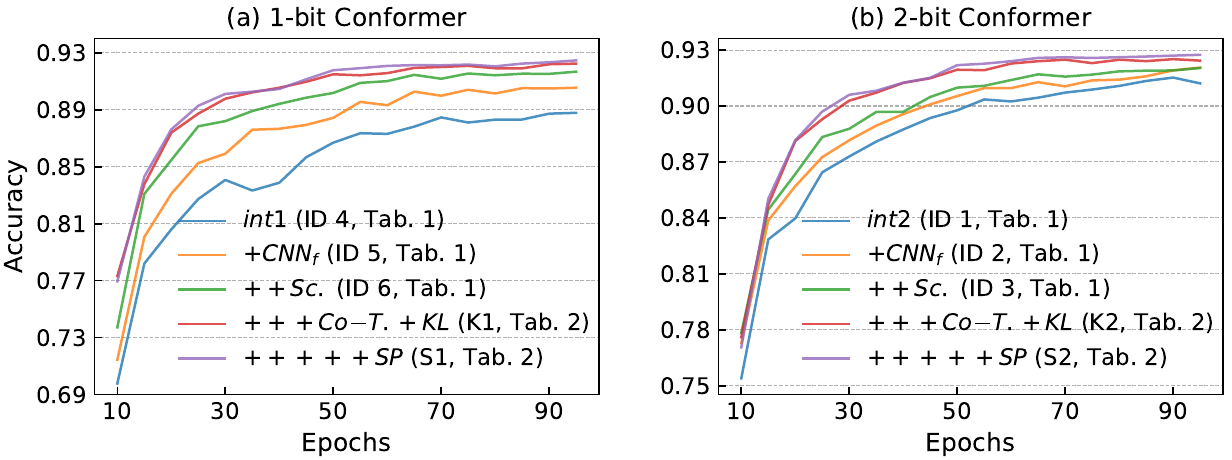}
    \vspace{-0.3cm}
    \caption{ Accuracy ($\uparrow$) of different 1-bit (a) and 2-bit (b) conformer systems on the validation set as a function of training epochs. ``\textit{int}1" and ``\textit{int}2" represent naively quantized 1-bit and 2-bit conformer systems, respectively. ``\textit{CNN}$_f$" denotes keeping the convolution modules with full-precision. ``\textit{Sc.}" denotes using learnable scaling factors. ``\textit{Co-T.}", ``\textit{KL}", and ``\textit{SP}" are for techniques of co-training, KL-regularization, and stochastic precision, respectively. The results are plot every 5 epochs when training. The following ID numbers denote the correspondence with the results in Tables~\ref{tab1} and \ref{tab2}.}
    \label{fig2}
    \vspace{-0.4cm}
\end{figure}

\begin{table}[ht]
    \centering
    \vspace{-0.2cm}
    \caption{WER($\downarrow$) of individually quantized 2-bit and 1-bit conformer systems on NIST Hub5’00, RT02, and RT03 test sets.}
    \setlength\tabcolsep{3pt}
    \vspace{-0.2cm}
    \resizebox{\linewidth}{!}{
    \begin{tabular}{c|c|c|c|ccccccc|c}
         \toprule
         \multirow{2}{*}{ID} & \multirow{2}{*}{\shortstack{Model\\Bitwidth}} & \multirow{2}{*}{\shortstack{CNN\\Bitwidth}}& \multirow{2}{*}{\shortstack{+\\Scaling}}&\multicolumn{2}{c}{Hub5’00} & \multicolumn{3}{|c|}{RT02} & \multicolumn{2}{c|}{RT03} & {\multirow{2}{*}{Avg.}} \\
         \cline{5-11}
         & & & & swbd & calhm & \multicolumn{1}{|c}{swbd1} & swbd2 & \multicolumn{1}{c|}{swbd3} & fsh & swbd & \\
         \midrule
         0&\multicolumn{3}{c|}{$float32$ Baseline}&7.4 & 15.2 & 8.9 & 13.0 & 15.9 & 10.6 & 16.6 & 12.86\\
         \midrule
         1&\multirow{3}{*}{2}&2&\xmark&9.0&17.3&10.5&14.4&18.2&12.1&19.1&14.74 \\
         2&&32&\xmark&8.1&16.0&9.4&14.1&17.1&11.2&17.5&13.68 \\
         3&&32&\cmark&7.8&15.9&9.4&13.2&16.7&10.8&17.2&13.33 \\
         \midrule
         4&\multirow{3}{*}{1}&1&\xmark&12.6&22.5&14.8&19.8&23.5&16.5&24.6&19.60 \\
         5&&32&\xmark&9.4&17.8&11.4&15.3&19.6&12.5&20.0&15.52 \\
         6&&32&\cmark&8.2&16.2&9.7&14.3&17.3&11.2&18.2&13.93 \\
         \bottomrule
    \end{tabular}
    }
    \label{tab1}
    \vspace{-0.5cm}
\end{table}

\begin{table}[ht]
    \centering
    \caption{WER($\downarrow$) of different Conformer systems on NIST Hub5’00, RT02, and RT03 test sets, where all the systems are by default trained under quantization co-training framework with learnable scaling factors. Systems with the same letter in their IDs are jointly trained and share weights (e.g., Sys.~O2 and Sys.~O1). A ``*" in the last column denotes no statistically significant (MAPSSWE~\cite{gillick1989some}, $\alpha$=0.05) WER increase is observed over the full-precision baseline system (ID 0). Performance-lossless compression results are highlighted with cell colors.}
    \setlength\tabcolsep{1pt}
    \vspace{-0.1cm}
    \resizebox{\linewidth}{!}{
    \begin{tabular}{l|c|c|c|c|c|ccccccc|c}
         \toprule
         \multirow{2}{*}{ID} & \multirow{2}{*}{\shortstack{Model\\Bitwidth}} &\multirow{2}{*}{+KL} &\multirow{2}{*}{\shortstack{+Stoc.\\\ \ \ Prec.}}&\multirow{2}{*}{\shortstack{CNN\\Bitwidth}}&\multirow{2}{*}{\shortstack{Comp.\\Ratio}}&\multicolumn{2}{c}{Hub5’00} & \multicolumn{3}{|c|}{RT02} & \multicolumn{2}{c|}{RT03} & {\multirow{2}{*}{Avg.}} \\
         \cline{7-13}
         & & & & & & swbd & calhm & \multicolumn{1}{|c}{swbd1} & swbd2 & \multicolumn{1}{c|}{swbd3} & fsh & swbd & \\
         \midrule
         0&\multicolumn{4}{c|}{$float32$ Baseline}&1.0x&7.4 & 15.2 & 8.9 & 13.0 & 15.9 & 10.6 & 16.6 & 12.86\\
         \midrule
         O2&\multirow{3}{*}{2}&\xmark&\xmark&\multirow{6}{*}{32}&\multirow{3}{*}{7.6x}&7.7&15.9&9.2&13.5&16.4&11.0&17.5&13.37 \\
         K2&&\cmark&\xmark&&&7.4&15.3&9.1&13.2&16.0&10.7&17.0&13.02 \\
         S2&&\cmark&\cmark&&&\cellcolor{orange!30}7.2&\cellcolor{orange!30}15.2&\cellcolor{orange!30}8.9&\cellcolor{orange!30}12.6&\cellcolor{orange!30}15.8&\cellcolor{orange!30}10.2&\cellcolor{orange!30}16.8&\cellcolor{orange!30}12.74$^*$ \\
         
         \cline{1-4}\cline{6-14}
         O1&\multirow{3}{*}{1}&\xmark&\xmark&&\multirow{3}{*}{9.7x}&8.0&16.4&9.9&13.9&17.3&11.8&18.1&13.98 \\
         K1&&\cmark&\xmark&&&7.8&15.7&9.2&13.2&16.5&10.9&17.3&13.33 \\
         S1&&\cmark&\cmark&&&\cellcolor{cyan!50}7.5&\cellcolor{cyan!50}15.6&\cellcolor{cyan!50}9.1&\cellcolor{cyan!50}13.1&\cellcolor{cyan!50}15.8&\cellcolor{cyan!50}10.5&\cellcolor{cyan!50}17.1&\cellcolor{cyan!50}12.99$^*$ \\
         \midrule
         E2&\multirow{3}{*}{2}&\multirow{9}{*}{\cmark}&\multirow{9}{*}{\cmark}&8&13.1x&7.2&15.0&8.6&12.8&16.0&10.6&16.8&12.81$^*$ \\
         F2&&&&4&14.9x&7.4&15.0&9.0&12.8&16.3&10.5&16.7&12.86$^*$ \\
         T2&&&&2&16.0x&7.6&15.5&9.2&13.2&16.2&10.9&17.5&13.26 \\
         \cline{1-2}\cline{5-14}
         E1&\multirow{3}{*}{1}&&&8&21.1x&7.6&15.4&9.1&13.2&16.1&10.6&17.1&13.06 \\
         F1&&&&4&26.2x&7.6&15.6&9.3&13.1&16.5&10.8&17.2&13.20 \\
         T1&&&&2&29.8x&7.9&16.2&9.4&13.6&17.0&11.2&17.9&13.69 \\
         \cline{1-2}\cline{5-14}
         E3&\multirow{3}{*}{1.5}&&&8&16.2x&\cellcolor{gray!40}7.6&\cellcolor{gray!40}15.4&\cellcolor{gray!40}8.9&\cellcolor{gray!40}13.2&\cellcolor{gray!40}16.0&\cellcolor{gray!40}10.7&\cellcolor{gray!40}16.9&\cellcolor{gray!40}12.99* \\
         F3&&&&4&19.0x&7.6&15.2&9.1&13.1&16.5&10.6&17.1&13.07 \\
         T3&&&&2&20.8x&7.8&15.8&9.5&13.4&16.8&11.2&17.7&13.55 \\
         \midrule
         1&\multicolumn{4}{c|}{\cite{li2023lossless} ($int2$ + CNN$_{int2}$)}&16.0x&8.1&16.3&9.5&13.9&17.2&11.1&17.8&13.77 \\
         \bottomrule
    \end{tabular}
    }
    \label{tab2}
    \vspace{-0.2cm}
\end{table}

\begin{table}[ht]
    \centering
    \caption{WER($\downarrow$) comparison with different quantization methods on the LibriSpeech dataset. A ``*" mark denotes no statistically significant (MAPSSWE~\cite{gillick1989some}, $\alpha$=0.05) WER increase is observed over the full-precision baseline system. A ``$^\Delta$" means our implementation. The best results are highlighted in bold.}
    \vspace{-0.2cm}
    \setlength\tabcolsep{5pt}
    \resizebox{\linewidth}{!}{
    \begin{tabular}{l|l|cccc|c|c|c}
        \toprule
         \multirow{2}{*}{ID}&\multirow{2}{*}{Model (Conformer)}&\multicolumn{2}{c}{test}&\multicolumn{2}{c|}{dev}&Model&\# of Extra & Comp.   \\
         \cline{3-6}
         &&clean&other&clean&other&Size&Quant. Param.&Ratio\\
         \midrule
         0&$float32$ Baseline&2.55&6.41&2.33&6.47&465MB&0&1.0x\\
         \midrule
         1&$int2$\cite{yuancompressed}&3.21&7.88&-&-&38MB&-&12.9x\\
         2&$int2$\cite{li2023lossless}&3.33&7.76&3.12&7.95&42MB&157K&10.6x\\
         F2&$int2$ (\textbf{Ours}, share weights with F1)&\textbf{2.61$^*$}&\textbf{6.30$^*$}&\textbf{2.45}&\textbf{6.42$^*$}&38MB&\textbf{204}&12.2x\\
         \midrule
         3&$int1$\cite{yuancompressed}&3.36&8.36&-&-&24MB&-&20.5x\\
         4&$int1$\cite{rybakov2024usm}$^\Delta$(absmean)&3.45&8.16&3.24&8.20&34MB&157K&13.1x\\
         F1&$int1$ (\textbf{Ours}, share weights with F2)&\textbf{2.79}&\textbf{6.44$^*$}&\textbf{2.57}&\textbf{6.56$^*$}&28MB&\textbf{204}&16.6x\\
         \midrule
         5&$int8$\cite{kim2022integer}&3.06&7.06&2.75&6.95&124MB&$>$157K&4.0x\\
         6&$int6$\cite{kim2022integer}&4.03&8.48&3.63&8.21&93MB&$>$157K&5.3x\\
         \bottomrule

    \end{tabular}
    }
    \label{tab3}
    \vspace{-0.6cm}
\end{table}

\vspace{-0.2cm}
\subsection{Results for Individually Quantized Conformer}
\vspace{-0.1cm}

We first separately perform 2-bit and 1-bit quantization of Conformer systems to study how to improve their own performance without the guidance or regularization of other systems. The results are reported in Table~\ref{tab1}. For a clearer comparison, we average the WER over the three test sets, shown in the last column of the Table~\ref{tab1}. There are several observations. \textbf{1)} Significant WER increases can be found for systems obtained with naive quantization-aware training (QAT). The 2-bit naively quantized system (ID~1) has a 14.6\% relative WER increase and the 1-bit system (ID~4) has a severer relative WER increase of 52.4\% against the full-precision baseline (ID~0). \textbf{2)} However, when keeping the convolution module full precision (CNNs that only have a minor contribution to the size) of the encoder, the WER has an obvious reduction for both 2-bit and 1-bit systems (IDs~2 and 5 vs. IDs~1 and 4). This is also consistent with the observations in previous studies~\cite{fasoli20214,rybakov20232,rybakov2024usm}. \textbf{3)} When using a learnable scaling factor, the WER can be further reduced for both 2-bit and 1-bit systems (IDs~3 and 6 vs. IDs~2 and 5).

\vspace{-0.1cm}
\subsection{Results for the Proposed Quantization Co-Training}
\vspace{-0.cm}
This section studies the efficacy of our proposed framework, namely quantization co-training with KL-regularization and stochastic precision. Through the above investigations in Table~\ref{tab1}, although the performance degradation for 2-bit and 1-bit Conformer systems are reduced from relative (absolute) WER increases of 14.62\% (1.88\%) and 52.41\% (6.74\%) to 3.65\% (0.47\%) and 8.32\% (1.07\%), respectively, they are far from a performance-lossless compression due to a statistically significant WER increase can still be observed. Hence, in addition to applying learnable scale factors, we further perform quantization with our proposed co-training framework, where the 2-bit and 1-bit models are jointly trained and share weights. The results are reported in Table~\ref{tab2}, where the WERs are also averaged for convenient comparisons. Remarkably, our approach achieved performance-lossless quantization for both 2-bit and 1-bit Conformer systems while incurring no statistically significant WER increase against the full-precision baseline (IDs~S2 and S1 vs. ID~0). Then, we demonstrate how our approach improves the performance of quantized models by progressively applying each of the proposed techniques. \textbf{1)} With the bare co-training method, the obtained 2-bit and 1-bit Conformer systems (IDs~O2 and O1 in Table~\ref{tab2}) have performance comparable to the individually quantized 2-bit and 1-bit ones (IDs~3 and 6 in Table~\ref{tab1}) while saving storage cost as the co-trained ones share weights. \textbf{2)} When adding the KL-divergence regularization to let the 2-bit model explicitly guide the 1-bit model during the co-training, the WERs are further reduced by 2.62\% (0.35\%) and 4.65\% (0.65\%) relative (absolute) for 2-bit and 1-bit sub-models (IDs~K2 and K1 vs. IDs~O2 and O1), respectively. \textbf{3)} By further using the stochastic precision method, the performance is improved again for the 2-bit and 1-bit sub-models with further relative (absolute) WERs reduction of 2.15\% (0.28\%) and 2.55\% (0.34\%) (IDs~S2 and S1 vs. IDs~K2 and K1), respectively. Fig.~\ref{fig2} shows the validation performance of the corresponding systems in Tables~1 and 2 during training, where obvious improvements can be consistently found by progressively adopting each of the proposed techniques.

To further increase the overall compression ratio, we investigate also quantizing the convolution modules. The results are reported in Table~\ref{tab2} from IDs~E2 to T3. Several trends can be observed. \textbf{1)} For 2-bit systems with CNNs no less than 4 bits, our method can still achieve performance-lossless compression (IDs~E2 and F2 vs. ID 0). \textbf{2)} For 1-bit systems, the impact of CNN quantization is more obvious than that of 2-bit systems, especially when CNNs are quantized with 2 bits. However, with the proposed methods, we still obtained a 1-bit system with an overall compression ratio of 26.2x at a cost of only 2.6\% relative (0.34\% absolute) WER increase over the full-precision baseline (ID~F1 vs. ID~0). In addition, our 1-bit system with 2-bit CNNs achieves a performance comparable to the entirely 2-bit system developed in previous work~\cite{li2023lossless} (ID~T1 vs. ID 1). \textbf{3)} As our stochastic precision method naturally enables various sub-models with bit-widths between 2 bits and 1 bit, we also evaluated some systems with half randomly sampled layers set to 1 bit and the rest kept with 2 bits, averaging to a mixed precision of 1.5 bits. We sample three 1.5-bit systems for each of the 8-, 4-, and 2-bit CNNs, and the average results are in Table~\ref{tab2} with IDs E3, F3 and T3. We achieve a performance-lossless quantization with a maximum overall compression ratio of 16.2 times (ID~E3 vs. ID~0)\footnote{The lossless compression of 1.5-bits systems is based on the observation that no less than 2 out of the 3 sampled systems are lossless.}.

\vspace{-0.1cm}
\subsection{Comparison with Other Methods}
\vspace{-0.1cm}
In this subsection, we conduct further experiments on the larger LibriSpeech dataset, in which we compare our method (i.e., taking the same setup as sys.~F2 and F1 in Table~\ref{tab2}) with state-of-the-art ASR quantization methods. The results are reported in Table~\ref{tab3}, where several trends can be found: \textbf{1)} Our method consistently and significantly outperforms other ASR quantization methods. For 2-bit quantization, our method achieved a reduction in relative (absolute) WER of up to 21.62\% (0.72\%) compared to other systems of similar model size (i.e., test-clean of ID~F2 vs ID~2 in Table~\ref{tab3}). Then, for the 1-bit case, our method achieved a reduction in relative (absolute) WER of up to 22.97\% (1.92\%) (i.e., test-other of ID~F1 vs ID~3 in Table~\ref{tab3}). \textbf{2)} Our method only requires negligible extra quantization parameters (i.e., 204), while other methods contain hundreds or thousands of times more extra parameters to perform quantization. \textbf{3)} Remarkably, our method achieves performance-lossless compression with a maximum compression ratio of 16.6 while incurring no statistically significant WER increase compared to the full-precision baseline system (i.e., test-clean, test-other, and dev-other of ID~F2 for 2-bit quantization and test-other and dev-other of ID~F1 for 1-bit quantization vs ID~0 in Table~\ref{tab3}). \textbf{4)} Furthermore, our method also contributes to storage savings because the obtained 1-bit and 2-bit systems share weights.

\vspace{-0.1cm}
\section{Conclusion}
\vspace{-0.cm}
We proposed a novel method of low-bit quantization based on the co-training framework with KL-divergence regularization and stochastic precision. We achieved performance-lossless 2-bit and 1-bit quantization of Conformer systems. We obtained a 16.2x compressed Conformer with no statistically significant WER increase and a 26.2x compressed Conformer with only minimal performance degradation on the 300-hr Switchboard corpus. Our method consistently showed superior performance on different datasets and outperformed state-of-the-art ASR quantization methods, while including only negligible quantization parameters.

\newpage

\section{Acknowledgements}
This research is supported by Hong Kong RGC GRF grant No. 14200220, 14200021, 14200324 and Innovation Technology Fund grant No. ITS/218/21.

\bibliographystyle{IEEEtran}
\bibliography{mybib}

\end{document}